\documentclass[english,twocolumn,prb,a4paper]{revtex4}
\usepackage{graphicx}
\usepackage{amssymb}

\usepackage{graphics}
\usepackage{babel}

\begin{document}

\title{Pomeranchuk and Topological Fermi Surface Instabilities from Central
Interactions}

\author{J. Quintanilla}

\affiliation{ISIS Spallation Facility, CCLRC Rutherford Appleton Laboratory, Chilton, Didcot,
OX11 0QX, U.K.}

\author{A. J. Schofield}

\affiliation{School of Physics and Astronomy, University of Birmingham, Edgbaston,
Birmingham, B15 2TT, U.K.}

\begin{abstract}
We address at the mean field level the emergence of a Pomeranchuk
instability in a uniform Fermi liquid with \emph{central} particle-particle
interactions. We find that Pomeranchuk instabilities with all symmetries except $l=1$
can take place if the interaction is repulsive and has a finite range
$r_{0}$ of the order of the inter-particle distance. We demonstrate
this by solving the mean field equations analytically for an explicit model interaction, as well as numerical results for more realistic potentials.
We find in addition to the Pomeranchuk instability other, subtler phase transitions in
which the Fermi surface changes topology without rotational symmetry-breaking.
We argue that such interaction-driven topological transitions
may be as generic to such systems as the Pomeranchuk instability. 
\end{abstract}

\maketitle

\section{Introduction}

Experimental evidence of {}``hidden'' phases of itinerant electron
systems \cite{tallon_99,kim_03,grigera_04} and the prospect of
realizing novel conditions in layered heterostructures and ultra-cold
gases have led to increased efforts to identify unconventional phase
transitions and predict their manifestations. To give three examples:
the Fulde-Ferrell-Larkin-Ovchinnikov state has been proposed in
organic superconductors \cite{balicas_01}, superconductor-ferromagnet
heterostructures \cite{annett_05} and in imbalanced mixtures of
ultracold atoms \cite{FFLO_in_atoms,FFLO_in_atoms-2}; a supersolid
phase is a possibility in Bose gases loaded on optical lattices
\cite{goral_02}; and a {}``$d$-density wave'' may be realized in ladder
compounds \cite{DDW-2,DDW-3} and possibly {}``hide'' in the phase diagram of cuprate
superconductors \cite{DDW}, where other hidden order parameters have been proposed \cite{varma1,varma2}. 

In this context there has been a surge of interest in the Pomeranchuk
Instability (PI) \cite{pomeranchuk_58}. Through it a Fermi liquid may
enter a {}``nematic'' state characterized by a deformed Fermi surface.
It has been argued that such an instability may take place in quantum
Hall systems \cite{QH,QH-2} and in the metamagnets
Sr$_{3}$Ru$_{2}$O$_{7}$ \cite{grigera_04} and URu$_{2}$Si$_{2}$
\cite{2005-Varma-Zhu}. Moreover there is evidence that the Hubbard
model has a phase with a distorted Fermi surface
\cite{HM,HM-2,HM-3,HM-4,HM-5,HM-6} and  the Emery model of a CuO$_2$
plane has been shown to have a nematic ground state in the strong
coupling limit  \cite{2004-Kivelson-Fradkin-Geballe}. 

More generally the PI is an interesting candidate unconventional phase
transition on account of its subtlety. Thus, considerable effort is
going into characterizing it theoretically
\cite{phase_diagram,phase_diagram-2,phase_diagram-3,collective,collective-2,collective-3,collective-4,collective-5} on the basis of phenomenological models
featuring anisotropic effective interactions. This approach is proving
very successful in establishing some generic features of the phase
diagram \cite{phase_diagram,phase_diagram-2,phase_diagram-3} and describing collective excitations and
quantum critical fluctuations \cite{collective,collective-2,collective-3,collective-4,collective-5}. On the other hand it
sidelines the question of how the anisotropy emerges in the first place
\cite{collective} and what other \cite{varma_05}, perhaps even subtler
instabilities may generically arise in such contexts. It is these
questions that we address here. 

\begin{figure}
\includegraphics[width=0.9\columnwidth,keepaspectratio]{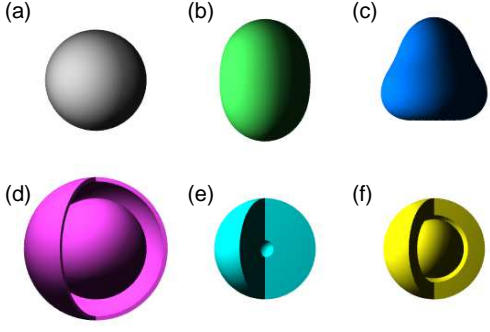}
\caption{\label{Fig.1} (color online) Different shapes and topologies of the Fermi surface.
From left to right, in the first row: (a) unpolarized, undeformed
Fermi sphere; (b) Fermi surface with an $l=2$ Pomeranchuk deformation;
(c) $l=3$. In the second row: (d) Fermi sphere surrounded by an additional
sheet of occupied states; (e) with a {}``hole'' of vacated states
at the center; and (f) with a shell of vacated states. }
\end{figure}

In this paper we present a mean field (MF) theory of the PI in a
three-dimensional, uniform fermion liquid with a \emph{central}
effective interaction potential $V(r)$. The authors of
Ref.~\onlinecite{collective} have pointed out that such interaction may
lead to a PI. Here we show that the emergence of the anisotropic state
from a Galilean invariant fluid 
requires repulsion with an intermediate range of the order of the
inter-particle distance. This is confirmed by explicit calculation for
a model interaction potential for which the theory can be solved
analytically. However we also find that the intermediate-range repulsion
leads, quite generally, to a different instability in which there is no
symmetry breaking but the topology of the Fermi surface changes. We discuss the
nature of this subtler quantum phase transition. 
A few instances of the two distinct types of Fermi surface shape 
instabilities that we find are pictured in Fig.~\ref{Fig.1}.
These two types of instability compete and we show that this conclusion
is robust when we consider more
more realistic finite range interactions.

\section{Mean field theory} 

To motivate a microscopic theory of the PI we start by recalling the
original, phenomenological theory due to Pomeranchuk \cite{pomeranchuk_58}.
Like him, we start with an unpolarized Fermi sphere and consider an
infinitesimal change of the occupation numbers, $N_{\mathbf{k},\sigma\sigma}\to N_{\mathbf{k},\sigma\sigma}+\delta N_{\mathbf{k},\sigma\sigma},$
arising from an angle-dependent modulation of the Fermi vector, $k_{F}\to k_{F}+\delta k_{F}\left(\theta\right)$,
$k_{F}\to k_{F}+\sigma\delta k_{F}\left(\theta\right)$ in the symmetric
or antisymmetric spin channel, respectively %
\footnote{We assume a $\mathbf{k}$-independent quantization axis for the spin
and choose this to be in the $z$ direction. One could envisage richer
states in which such axis cannot be defined, for example analogous
to the Balian-Werthamer superconducting state \cite{leggett_75}. A state 
of this type has been discussed in Ref.~\onlinecite{2004-Wu-Zhang}.}. We then use Landau's expression for the corresponding 
change in the ground state energy: $E\to E+\delta E,$ with\begin{eqnarray}
\delta E & = & \sum_{\mathbf{k}}\varepsilon\left(\mathbf{k}\right)\delta N_{\mathbf{k}}\label{0.1}\\
 &  & +\frac{1}{2}\sum_{\mathbf{k},\mathbf{k}'}\left\{ f^{s}\left(\mathbf{k},\mathbf{k}'\right)\delta N_{\mathbf{k}}\delta N_{\mathbf{k}'}+f^{a}\left(\mathbf{k},\mathbf{k}'\right)\delta\mathbf{S}_{\mathbf{k}} \cdot \delta\mathbf{S}_{\mathbf{k}'}\right\} .\nonumber \end{eqnarray}
Here $\delta N_{\mathbf{k}}=\sum_{\sigma}\delta N_{\mathbf{k},\sigma\sigma}$
and $\delta S_{\mathbf{k}}^{i}=\frac{1}{2}\sum_{\sigma,\gamma}\sigma_{\sigma\gamma}^{i}\delta N_{\mathbf{k},\gamma\sigma}.$
Requiring $\delta E<0$ leads to the PI conditions\begin{equation}
1+F_{l}^{a,s}/\left(2l+1\right)<0,\label{0.2}\end{equation}
in terms of the Landau parameters, defined by\footnote{We follow the
notations of Ref.~\onlinecite{leggett_75}.} 
\begin{equation}
f^{s,a}\left(k_F\hat{\mathbf{k}},k_F\hat{\mathbf{k}}'\right)=\frac{\pi^{2}\hbar
v_{F}}{\Omega k_{F}^{2}}\sum_{l=0}^{\infty}F_{l}^{s,a}P_{l}\left(\hat{\mathbf{k}}.\hat{\mathbf{k}}'\right),\label{0.3}
\end{equation}
 where $k_{F}$ is the radius of the Fermi sphere, $v_{F}$ is the
Fermi velocity and $P_{l}\left(x\right)$ is the $l^{{\rm th}}$ Legendre
polynomial. For $l=0,$ Eq.~(\ref{0.2}) describes a quantum gas-liquid
transition (in the symmetric spin channel, $s$) or a Stoner instability
(in the antisymmetric channel, $a$). For $l>0$ it describes a Pomeranchuk
instability.

One crucial aspect of Pomeranchuk's theory is that it describes the
instability in terms of the phenomenological Landau parameters, $F_{l}^{s,a}$.
Here we want to establish the mechanism whereby the PI could take
place in a system with a given microscopic Hamiltonian of the form\begin{eqnarray}
H & = & \int d^{3}\mathbf{r}\sum_{\sigma}\hat{c}_{\mathbf{r},\sigma}^{+}\left[\frac{1}{2m}\left(\frac{\hbar}{i}\nabla\right)^{2}-\mu\right]\hat{c}_{\mathbf{r},\sigma}\label{1.1}\\
 &  & +\frac{1}{2}\sum_{\sigma,\sigma'}\int d^{3}\mathbf{r}\int d^{3}\mathbf{r}'\hat{c}_{\mathbf{r},\sigma}^{+}\hat{c}_{\mathbf{r}',\sigma'}^{+}V\left(\left|\mathbf{r}-\mathbf{r}'\right|\right)\hat{c}_{\mathbf{r}',\sigma'}\hat{c}_{\mathbf{r},\sigma},\nonumber \end{eqnarray}
where $V\left(\left|\mathbf{r}-\mathbf{r}'\right|\right)$ is a local,
non-retarded, spin-independent and \emph{central} interaction potential. 
We address this question using MF theory. Our MF Hamiltonian is \begin{equation}
H_{0}=\sum_{\mathbf{k},\sigma}\varepsilon_{\sigma}\left(\mathbf{k}\right)\hat{c}_{\mathbf{k},\sigma}^{\dagger}\hat{c}_{\mathbf{k},\sigma},\label{1.2}\end{equation}
where $\hat{c}_{\mathbf{k},\sigma}^{+}\equiv\Omega^{-1/2}\int d^{D}\mathbf{r}e^{-i\mathbf{k}.\mathbf{r}}\hat{c}_{\mathbf{r},\sigma}^{+}.$
This describes independent electrons with an \emph{arbitrary} dispersion
relation $\varepsilon_{\sigma}\left(\mathbf{k}\right),$ which we treat
as our variational parameter. Note that this MF couples only to the
occupation number in $\mathbf{k}$-space, $\hat{N}_{\mathbf{k},\sigma\sigma}=\hat{c}_{\mathbf{k},\sigma}^{\dagger}\hat{c}_{\mathbf{k},\sigma}$.
Our theory thus preserves translational and gauge symmetry, but it
can nevertheless break rotational symmetry if the dispersion relation
becomes anisotropic. For example, a nematic Fermi liquid state may be entered
through a PI. It is an example of an ``electronic liquid crystal state'' \cite{liquid_crystal}.

Although our main results refer to the ground state, the derivation of the basic equations of the theory is much simpler at finite temperature. We thus approximate the free energy by $F\approx\left\langle H-H_{0}\right\rangle _{0}+F_{0}$,
where $\left\langle \ldots\right\rangle _{0}=Z_{0}^{-1}{\rm Tr}\left\{ e^{-\beta H_{0}}\ldots\right\} $
with $Z_{0}={\rm Tr}\left\{ e^{-\beta H_{0}}\right\} $ and $F_{0}=-\beta^{-1}\ln Z_{0}$.
It takes the form
\begin{eqnarray}
F & = & \sum_{\mathbf{k},\sigma}\left\{ \begin{array}{c}
\\\\\\\end{array}\right.\hspace{-10pt}N_{\mathbf{k},\sigma}\left[\begin{array}{c}
\\\\\end{array}\right.\hspace{-10pt}-\frac{1}{2\Omega}\sum_{\mathbf{k}'}V\left(\left|\mathbf{k}-\mathbf{k}'\right|\right)N_{\mathbf{k}',\sigma}\label{1.3}\\
 &  & +\frac{\bar{V}}{2\Omega}\sum_{\mathbf{k}',\sigma'}N_{\mathbf{k}',\sigma'}+\frac{\hbar^{2}\left|\mathbf{k}\right|^{2}}{2m}-\mu-\varepsilon_{\sigma}\left(\mathbf{k}\right)\hspace{-10pt}\left.\begin{array}{c}
\\\\\end{array}\right]\nonumber \\
 &  & -\frac{1}{\beta}\ln\left[1+e^{-\beta\varepsilon_{\sigma}\left(\mathbf{k}\right)}\right]\hspace{-10pt}\left.\begin{array}{c}
\\\\\\\end{array}\right\} \nonumber \end{eqnarray}
where $\bar{V}=\int d^{3}\mathbf{R}V\left(\left|\mathbf{R}\right|\right)$
is the uniform component of the interaction potential and $V\left(\mathbf{K}\right)=\int d^{3}\mathbf{R}e^{-i\mathbf{K}.\mathbf{R}}V\left(\left|\mathbf{R}\right|\right)$
its Fourier transform. The occupation numbers in $\mathbf{k}$-space
are given by 
\begin{equation}
N_{\mathbf{k},s}=\left[1+e^{\beta\varepsilon_{\sigma}\left(\mathbf{k}\right)}\right]^{-1}.
\label{1.4}
\end{equation}
Requiring that $F$ be stationary yields 
\begin{equation}
\varepsilon_{\sigma}\left(\mathbf{k}\right)=\frac{\hbar^{2}\left|\mathbf{k}\right|^{2}}{2m}-\mu+\frac{1}{\Omega}\sum_{\mathbf{k}'\sigma'}\left\{ \bar{V}-\delta_{\sigma,\sigma'}V\left(\left|\mathbf{k}-\mathbf{k}'\right|\right)\right\} N_{\mathbf{k}',\sigma'}.
\label{1.5}
\end{equation}
In the low-temperature limit, $\beta\to\infty,$ Eqs.~(\ref{1.3},\ref{1.4})
become \begin{eqnarray}
E & = & \sum_{\mathbf{k},\sigma}N_{\mathbf{k},\sigma}\left[\begin{array}{c}
\\\\\end{array}\right.\hspace{-10pt}-\frac{1}{2\Omega}\sum_{\mathbf{k}'}V\left(\left|\mathbf{k}-\mathbf{k}'\right|\right)N_{\mathbf{k}',\sigma}\label{1.3'}\\
 &  & +\frac{\bar{V}}{2\Omega}\sum_{\mathbf{k}',\sigma'}N_{\mathbf{k}',\sigma'}+\frac{\hbar^{2}\left|\mathbf{k}\right|^{2}}{2m}-\mu\hspace{-10pt}\left.\begin{array}{c}
\\\\\end{array}\right];\nonumber \\
N_{\mathbf{k},\sigma} & = & \Theta\left[-\varepsilon_{\sigma}\left(\mathbf{k}\right)\right];\label{1.4'}\end{eqnarray}
and our MF theory is equivalent to trying variationally the following
ground state: \begin{equation}
\left|\Psi\right\rangle =\prod_{\varepsilon_{\sigma}\left(\mathbf{k}\right)<0}\hat{c}_{\mathbf{k},\sigma}^{\dagger}\left|0\right\rangle .\label{1.2'}\end{equation} 

Eq.~(\ref{1.3'}) is our MF approximation to the ground state energy.
Variation with respect to the occupation numbers yields an expression
identical to Eq.~(\ref{0.1}), except that the phenomenological functions
$\varepsilon_{\sigma}\left(\mathbf{k}\right)$ (which here may depend
on the spin) and $f^{s,a}\left(\mathbf{k},\mathbf{k}'\right)$ are
now derived from our microscopic parameters via Eq.~(\ref{1.5})
and \begin{equation}
f^{s,a}\left(\mathbf{k},\mathbf{k}'\right)=\frac{1}{\Omega}\left\{
\eta\bar{V}-\frac{1}{2}V\left(|\mathbf{k}-\mathbf{k}'|\right)\right\} ,\label{1.6}\end{equation}
 where $\eta=1,0$ in the $s,\, a$ channels, respectively.

Let us pause briefly to note the following subtlety. Eqs.~(\ref{1.3'},\ref{1.4'}), from which all the subsequent results follow, could have been derived by minimizing the energy of the trial state given in Eq.~(\ref{1.2'}). However note this only determines the Fermi surface, but it under-determines the dispersion relation $\varepsilon_{\sigma}({\bf k})$. The justification of the particular form given in Eq.~(\ref{1.5}) thus relies on the assumption that the low-lying excitations correspond to re-arrangements of the electrons in momentum space, whose energy is given by Eq.~(\ref{0.1}) [or, equivalently, that the equilibrium state at finite temperatures can be adequately described by the mean field Hamiltonian of Eq.~(\ref{1.2}).] This is necessary to justify the language we use below, e.g. in defining the Fermi velocity in terms of $\varepsilon_{\sigma}({\bf k})$. We stress, however, that the results themselves refer only to the equilibrium shape of the Fermi surface in the !
 ground state and are therefore more general, and independent of the meaning assigned to $\varepsilon_{\sigma}({\bf k})$.

To study the PI in our microscopic model we postulate an unpolarized,
spherical Fermi surface\begin{equation}
N_{\mathbf{k},\sigma}=\Theta\left(k_{F}-\left|\mathbf{k}\right|\right),\label{2.1}\end{equation}
 completely described by the Fermi vector $k_{F}>0$ {[}Fig.~\ref{Fig.1}
(a){]}, and use the above equations to determine whether the system
has a PI. In the state described by Eq.~(\ref{2.1}), the electron
dispersion relation of Eq.~(\ref{1.5}) is given by \begin{eqnarray}
\varepsilon\left(\left|\mathbf{k}\right|\right) & = & \frac{\hbar^{2}}{2m}\left(\left|\mathbf{k}\right|^{2}-k_{F}^{2}\right)-\frac{2k_{F}^{2}}{\pi}\int_{0}^{\infty}dr\, r\, V\left(r\right)\label{2.2}\\
 &  & j_{1}\left(k_{F}r\right)\left[j_{0}\left(\left|\mathbf{k}\right|r\right)-j_{0}\left(k_{F}r\right)\right].\nonumber \end{eqnarray}
This yields the following expression for the Fermi velocity:\begin{eqnarray}
v_{F} & = & \frac{\hbar}{m}k_{F}+\frac{2k_{F}^{2}}{\hbar\pi}\int_{0}^{\infty}dr\, r^{2}\, V\left(r\right)\, j_{1}\left(k_{F}r\right)^{2}\label{2.3}\\
 & = & \frac{\hbar}{m}k_{F}+\frac{p_{F}^{2}}{\left(2\pi\hbar\right)^{3}}V_{1},\nonumber \end{eqnarray}
where in the second line we have expressed $v_{F}$ in terms of one
of the couping constants defined by Eq.~(\ref{2.5}), below. Note
that the state described by Eq.~(\ref{2.1}) requires $v_{F}>0$.
Together Eqs.~(\ref{1.6}) and (\ref{2.3}) give the Landau parameters
in Eq.~(\ref{0.3}),
\begin{equation}
F_l^{a,s}=\frac{2l+1}{(2\pi)^3}\frac{k_F^2}{\hbar v_F}\left(\eta
\delta_{l,0}8\pi
\bar{V}-V_l\right),
\label{Eq.16}
\end{equation} 
in terms of the microscopic parameters of the
model.
This, in turn, allows us to express the PI equations as \begin{equation}
V_{l}-\eta\delta_{l,0}8\pi\bar{V}>\frac{\left(2\pi\hbar\right)^{3}}{p_{F}^{2}}v_{F},\,\,\left[=\frac{\left(2\pi\hbar\right)^{3}}{mp_{F}}+V_{1}\right]\label{2.4}\end{equation}
where the strength of the interaction potential in a given angular
momentum channel $l=0,1,2,\ldots$ is given by \begin{equation}
V_{l}=\left(4\pi\right)^{2}\int_{0}^{\infty}dr\, r^{2}\, V\left(r\right)\, j_{l}\left(k_{F}r\right)^{2}.\label{2.5}\end{equation}
For antisymmetric instabilities, $\eta=0$, Eq.~(\ref{2.4}) takes the following, explicit form:
\begin{equation} 
   \int_{0}^{\infty}dr \, 4\pi r^{2} V\left(r\right)    
\left[j_{l}\left(k_{F}r\right)^{2}-j_{1}\left(k_{F}r\right)^{2}
\right]>\frac{2\pi^2\hbar^2}{mk_{F}}.
\end{equation}

Eq.~(\ref{2.4}) is our microscopic expression of the PI condition
of Eq.~(\ref{0.2}). It is valid, within our MF ansatz of Eq.~(\ref{1.2}),
for any system whose Hamiltonian has the form given by Eq.~(\ref{1.1}).
From it we can derive a series of conclusions concerning a Pomeranchuk
instability in an isotropic system with central interactions:

\begin{enumerate}
\item \label{enu:1}Purely attractive interactions can only lead to the
gas-liquid transition ($l=0,\eta=1$) %
\footnote{Our MF Hamiltonian $H_{0}$ lacks any terms hybridizing particles
with holes and therefore cannot describe the key instability present
for purely attractive interactions, namely superconductivity. When
this is included, all other instabilities are precluded \cite{nozieres_85,nozieres_92}. %
}; conversely, purely repulsive interactions can only lead to the
Stoner or PI.
\item There are no $l=1$ Pomeranchuk instabilities. This is the type of
PI \cite{varma_05,2005-Varma-Zhu,2004-Wu-Zhang} where rotational symmetry-breaking
is achieved by displacing the Fermi surface so as to set up a charge ($s$) or spin ($a$) current. This
is quite a general consequence of the well known relation between the
effective mass and the Landau parameter, $F_1^s$, in a Galilean invariant system, which is captured by Eqs.~(\ref{2.3}) and (\ref{Eq.16}). On
the other hand for spin-dependent interactions (or in lattice systems), 
not considered here, we may
have $F_1^s \neq F_1^a$ and then the instabilities considered in 
Refs.~\cite{varma_05,2005-Varma-Zhu,2004-Wu-Zhang} could be realized.
\item The PI for $l\geq2$ is degenerate in the spin channel. These instabilities
break rotational symmetry by changing the shape of the Fermi surface,
without generating any currents of charge or spin ---see Fig.~\ref{Fig.1}(b,c).
Our result implies that, at the instability, it does not matter whether
the lobes of the spin-up and spin-down Fermi surface point in the
same direction %
\footnote{Inside the region of instability, higher order terms in the free energy
decide between these configurations.%
}. Note this is quite different from the situation at $l=0$ (see point
\ref{enu:1}, above). 
\item \label{enu:4}Finally, from Eq.~(\ref{2.5}) we can also deduce that
$V_{l}-V_{1}$ cannot be large and positive, as required by Eq.~(\ref{2.4}),
if the repulsive part of the interaction is of very short range $r_{0}\ll k_{F}^{-1}.$
In effect, $r^{2}j_{l}\left(k_{F}r\right)^{2}\sim r^{2\left(1+l\right)}$
for $r\ll k_{F}^{-1}$ so for such short-ranged interactions Eq.~(\ref{2.5})
gives $V_l \sim \int_0^{r_0} dr r^{2(l+1)} \sim r_0^{2l+3}$ whence for small $r_0$ Eq.~(\ref{2.4}) can only be satisfied for $l=0$. The extreme case of this is the repulsive contact potential $V\left(\left|\mathbf{r}\right|\right)=\left|u\right|\delta^{\left(3\right)}\left(\mathbf{r}\right),$
for which $r_{0}=0$ and $V_{l}=4\pi\left|u\right|\delta_{l,0}$.
For this potential, our theory leads only to the Stoner instability. 
\end{enumerate}

\section{Delta-shell model}

\begin{figure}
\includegraphics[%
  bb=50bp 50bp 266bp 201bp,
  width=1.0\columnwidth]{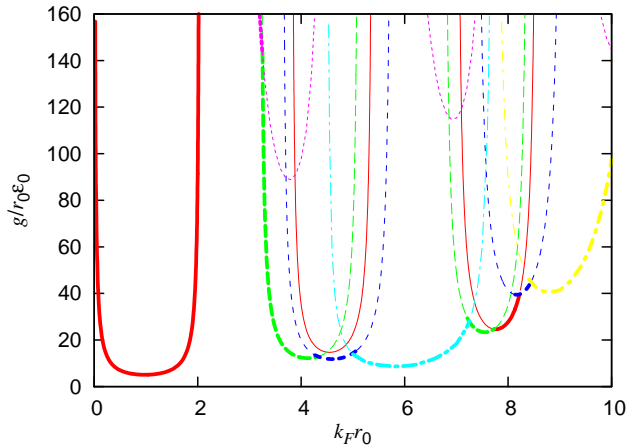}

\caption{\label{Fig.2} (color online) Phase diagram showing zero-temperature instabilities of the types depicted in Fig.~\ref{Fig.1} for the repulsive delta-shell model. The Stoner instability occurs above the solid line. The other instabilities described in Fig.~\ref{Fig.1} occur above the long dash line (b),
medium dash line (c), short dash line (d), long dash-dot line (e) and short
dash-dot line (f). The colors correspond with Fig.\ref{Fig.1}}
\end{figure}

From point \ref{enu:4} we conclude that repulsive interactions with
range at least of the order of the Fermi wavelength, $r_{0}\gtrsim k_{F}^{-1},$
are necessary for the PI in an isotropic system with isotropic interactions. 
We investigate this further by choosing
a specific form of the central interaction potential, namely the {}``delta-shell''
potential:\begin{equation}
V\left(\left|\mathbf{r}\right|\right)=g\delta^{\left(1\right)}\left(\left|\mathbf{r}\right|-r_{0}\right).\label{3.1}\end{equation}
This is an idealization of an interaction with a very sharp peak at
a particular inter-particle distance, $\left|\mathbf{r}\right|=r_{0}.$
The ``coupling constant'' $g$ has dimensions of energy $\times$ length and represents the product of the height and width of the potential barrier. 

For $g<0,$ this interaction potential can lead to superconductivity with unconventional pairing
\cite{2000-Quintanilla-Gyorffy,quintanilla_02}. Likewise, we expect that for $g>0$ it will
lead to a PI. Indeed Eq.~(\ref{2.5}) gives \begin{equation}
V_{l}=\left(4\pi\right)^{2}gr_{0}^{2}j_{l}\left(k_{F}r_{0}\right)^{2}\label{3.2}\end{equation}
so, depending on the value of $k_{F}r_{0},$ any value of $l$ may
become dominant. 

The particularly simple form of the interaction potential in Eq.~(\ref{3.1})
allows us to write the key expressions in our theory of the PI analytically.
In particular Eq.~(\ref{2.3}) giving the Fermi velocity on the Fermi
sphere reads \begin{equation}
v_{F}=\frac{\hbar}{m}k_{F}+g\frac{2\left(k_{F}r_{0}\right)^{2}}{\hbar\pi}j_{1}\left(k_{F}r_{0}\right)^{2}.\label{4.1}\end{equation}
 Together with Eq.~(\ref{3.2}) and $\bar{V}=g4\pi r_{0}^{2}$ these
equations allow us to write the following, simple form of the Stoner
and PI equations: \begin{equation}
j_{l}\left(k_{F}r_{0}\right)^{2}-j_{1}\left(k_{F}r_{0}\right)^{2}>\frac{\hbar^{2}}{2mr_{0}^{2}}\frac{\pi}{gk_{F}},\, l=0,1,2,\;\ldots\label{4.2}\end{equation}
If the interaction is very strong, $gk_{F}\gg\hbar^{2}/2mr_{0}^{2},$
this gives a sequence of fixed phase boundaries at $j_{l}\left(k_{F}r_{0}\right)=\pm j_{1}\left(k_{F}r_{0}\right).$
In the opposite limit of very weak interaction the unpolarized Fermi
sphere is, as expected, stable.

The solid, long dash and medium dash lines of Fig.~\ref{Fig.2} are the phase diagram obtained
by solving Eq.~(\ref{4.2}) for $l\leq3$. Note that there are effectively
only two, dimensionless parameters in the theory \cite{2000-Quintanilla-Gyorffy,quintanilla_02}: the ``effective range'' $k_{F}r_{0}$ and ``coupling constant'' $g/r_{0}\varepsilon_{0}$
(where $\varepsilon_{0}=\hbar^{2}/2mr_{0}^{2}$). The first of these parameters is the range of the interaction measured in units of $1/k_F$. The second is the product of the width of the potential barrier measured in units of that range and its height measured in units of the corresponding ``localization energy'' $\varepsilon_{0}$. As expected, for small
range, $k_{F}r_{0}\lesssim3,$ we only find the Stoner instability.
For longer ranges or, equivalently, higher densities, and sufficiently
large values of the dimensionless coupling constant (which we note that depends not only on $g$ but also on $m$ and $r_0$), the PI can take place. 

\begin{figure}
\includegraphics{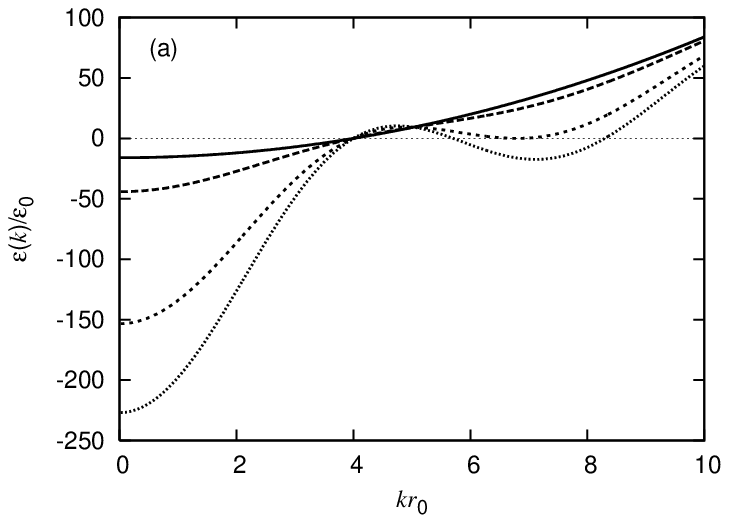}\\
\includegraphics{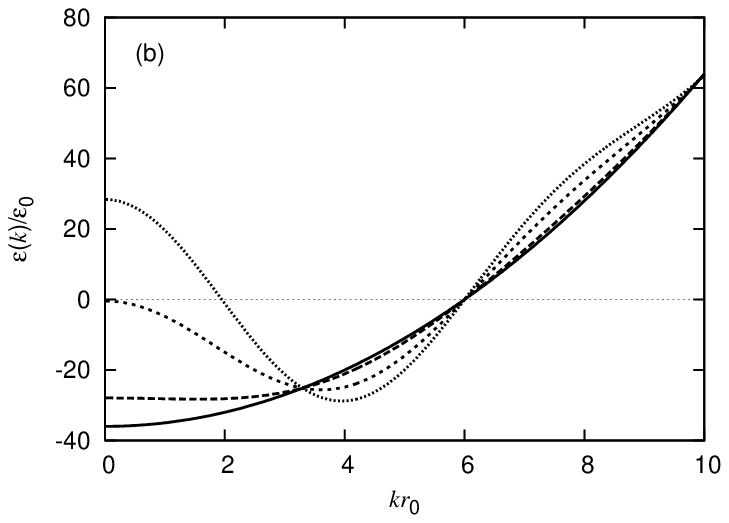}\\
\includegraphics{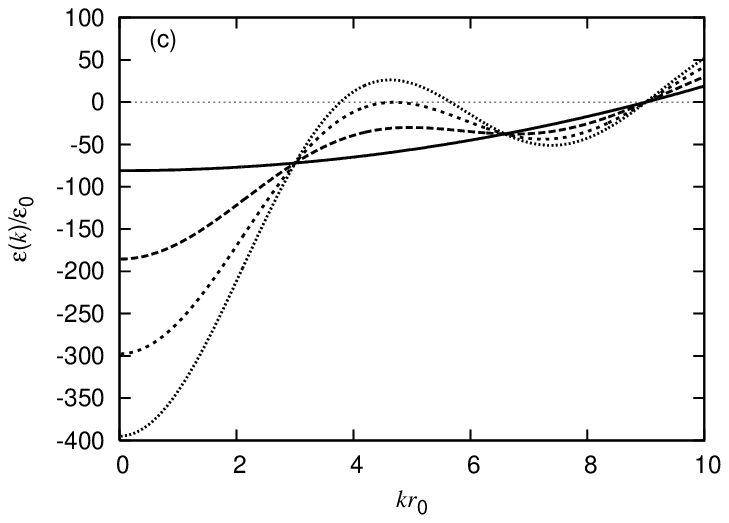}
\caption{\label{Fig.3}Electron dispersion relation for the {}``delta-shell''
model in the state with an unpolarized Fermi sphere: (a) $k_{F}r_{0}=4$,
$g/r_{0}\varepsilon_{0}=0\,(\textrm{solid}),\,20\,\textrm{(long
dash)},\,97.66\,\textrm{(short dash)},\,150\,\textrm{(dotted)}$;
(b) $k_{F}r_{0}=6$, $g/r_{0}\varepsilon_{0}=0,\,2,\,8.84,\,16$ (same
order); (c) $k_{F}r_{0}=9$, $g/r_{0}\varepsilon_{0}=0,\,20,\,41.40,\,60$.}
\end{figure}

\section{Topological transitions}

In addition to the anticipated PI, our analytic treatment of the
delta-shell model also reveals a competing class of Fermi surface
instabilities which, unlike the PI, occur without symmetry breaking.
Consider the electron dispersion relation in the
isotropic state, Eq.~(\ref{2.2}). For the delta-shell potential
it takes the form \begin{eqnarray}
\varepsilon\left(\left|\mathbf{k}\right|\right) & = & \frac{\hbar^{2}}{2m}\left(\left|\mathbf{k}\right|^{2}-k_{F}^{2}\right)-\frac{g}{r_{0}}\frac{2k_{F}^{2}r_{0}^{2}}{\pi}j_{1}\left(k_{F}r_{0}\right)\label{5.1}\\
 &  & \left[j_{0}\left(\left|\mathbf{k}\right|r_{0}\right)-j_{0}\left(k_{F}r_{0}\right)\right].\nonumber \end{eqnarray}
This is plotted in Fig.~\ref{Fig.3} for three different values of
$k_{F}r_{0}$. The free-electron dispersion relation is modified by
an oscillatory term due to electron-electron interactions. The period
of the oscillations is $\sim r_{0}^{-1}$. For small $g/r_{0}\varepsilon_{0}$
the effect of these is the usual renormalization of the effective
mass $m^{*}=p_{F}/v_{F}$, which follows from Eq.~(\ref{4.1}). However
at large $g/r_{0}\varepsilon_{0}$ the effect of interaction on this
{}``bare'' dispersion relation cannot be described simply as a renormalization
of $m$. In fact it can lead to a dramatic change of the state of
the system as the amplitude of the oscillations becomes large enough
that either 
\begin{enumerate}
\item \label{topo-a}the dispersion relation dips below the Fermi level somewhere outside
the Fermi sphere {[}Fig.~\ref{Fig.3}~(a){]},
\item \label{topo-b}it goes above the Fermi level at the center of the Fermi sphere {[}Fig.~\ref{Fig.3}~(b){]}
or
\item \label{topo-c}it peaks above the Fermi level at some intermediate $k$, $0<k<k_{F}$
{[}Fig.~\ref{Fig.3}~(c){]}.
\end{enumerate}
In either case, Eq.~(\ref{1.4'}) no longer reduces to Eq.~(\ref{2.1}).
Instead, either a thin shell of occupied states forms outside the
Fermi sphere {[}Fig.~\ref{Fig.1}~(d){]}, or states inside the Fermi
sphere become vacated {[}Fig.~\ref{Fig.1}~(e,f){]}. The associated
instabilities are quite distinct from the Stoner and PI, as the change
$\delta N_{\mathbf{k},\sigma},$ although infinitesimal, takes place
away from the Fermi surface. Instead, they are continuous phase transitions
in which no symmetry is broken, but the topology of the Fermi surface
changes. In that sense they are more reminiscent of the Lifshitz transition
\cite{lifshitz,lifshitz-2}. There is, however, a crucial difference,
namely that the present phase transitions are driven by electron-electron
interactions, which induce the fermions to {}``migrate'' to other
regions of reciprocal space, rather than by the band structure. On the basis 
of this one would expect the present instabilities to have a much stronger thermodynamic signature. 
For example, the transitions illustrated in Fig.~\ref{Fig.1}(d) and
(f), with their underlying dispersions of Fig.~\ref{Fig.3}(a) and (c)
respectively, result in the appearance of entire new Fermi surface
sheets with finite $k_F$. This will lead in mean-field theory to a
discontinuous jump in the density of states and hence in $C/T$. 

\begin{figure}
\includegraphics{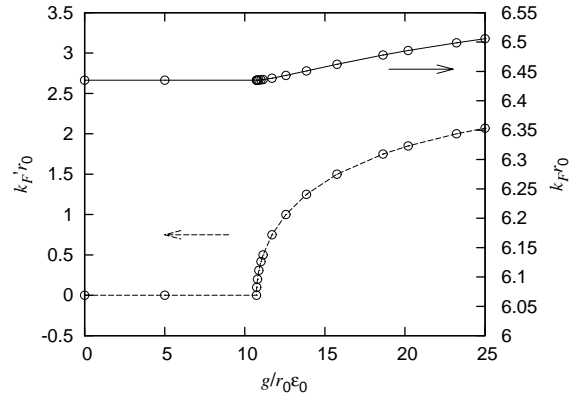}
\caption{\label{Fig.3bis}Evolution of the Fermi vector as a function of the dimensionless coupling constant $g/r_0\varepsilon_0$ for fixed particle density $N/\Omega=9 ~r_0^{-3}$. The plot shows both the ``original'' Fermi vector $k_F$ and the ``emerging'' Fermi vector $k_F'$, namely the radius of the sphere of empty states depicted in Fig.~\ref{Fig.1}~(e), as the system enters from below the corresponding dome in the phase diagram (see Fig.~\ref{Fig.2}).}
\end{figure}

A general framework to understand such topological quantum phase transitions has been put forward in Ref.~\onlinecite{2006-Volovik}. In this formalism, the Fermi surface is a vortex ``loop'' in a four-dimensional space and the phase transitions we have just described correspond to the nucleation of new loops. 
One can also view these instabilities as generalizations to dimension larger than one of the phenomenon of ``quantum Hall edge reconstruction''. The latter can be described as the emergence, due to interactions, of new Fermi points in the one-dimensional chiral Fermi liquid on the edge of a quantum Hall system \cite{1993-MacDonald-Yang-Johnson,1994-Chamon-Wen,2002-Wan-Yang-Rezayi,2006-CastroNeto-Guinea-Peres}. Indeed  in the instabilities described here always one of the new Fermi surfaces has negative Fermi velocity - analogous to the creation of left-moving quasiparticles in a right-moving chiral Fermi liquid \footnote{A similar phase transition has also been described in S. A. Artamonov,
V. R. Shaginyan and Yu. G. Pogorelov, \emph{JETP Lett.} \textbf{68},
942 (1998). It has also been suggested that the opening of a full
gap to single-particle excitations on approach to the Bose-Einstein
condensation limit in a $d$-wave superconductor should be regarded
as an interaction-driven Lifshitz transition: S. S. Botelho and C.
A. R. S\'{a} de Melo, \emph{Phys. Rev. B.} \textbf{71}, 134507 (2005).
A change of topology of the Fermi surface has also been noticed in
the RVB state on a frustrated square lattice {[}B.J. Powell, \emph{private
communication}{]}, for fermions on a lattice with boson-controlled
hopping {[}D.M. Edwards, Physica B {\bf 378-380}, 133-134 (2006) {]} and in the pseudogap region of the cuprate phase diagram [C.~M.~Varma and Lijun Zhu, cond-mat/0607777].}. 
Yang and Sachdev \cite{2006-Yang-Sachdev} have recently described the quantum critical fluctuations for a phase transition of type \ref{topo-a}, above [Figs.~\ref{Fig.1}~(d) and ~\ref{Fig.3}~(a)].

It is important to note that the plots in Fig.~\ref{Fig.3} correspond to evaluating Eq.~(\ref{5.1}) at a fixed value of $k_F$. For a fixed number of particles, $N$, such solutions are valid only {\em up to} the instability, as beyond it they would violate Luttinger's theorem. To describe the migration of electrons in reciprocal space mentioned above, which happens {\em beyond} the instability, it is necessary to determine $k_F$ self-consistently by requiring that the total number of particles be fixed. For example, for an instability of the type \ref{topo-b}, above [Figs.~\ref{Fig.3}~(b) and \ref{Fig.1}~(e)], this means that 
\begin{equation} 
   \frac{N}{\Omega}=2\frac{1}{(2\pi)^{3}}\frac{4\pi}{3}\left(k_F^3-k_F'^3\right),
\end{equation}
where $k_F'$ is the ``emerging'' Fermi vector at the centerer of the Fermi sphere, determined by $\varepsilon(k_F')=0$. This is demonstrated in Fig.~\ref{Fig.3bis}. Note that the way $k_F'$ grows as a function of coupling suggests thinking of this quantity as a sort of topological ``order parameter''. 

It is evident from Fig.~\ref{Fig.3} that the new sheet of electron-like
or hole-like Fermi surfaces are initially formed by localized states,
with $v_{F}=0.$ As we progress into the new state, $v_{F}$ becomes
finite. Conversely, if we run the process backward, the effective
masses on the additional Fermi surfaces diverge (except for the hole-like
Fermi surface in Fig.~\ref{Fig.1}~(e), for which the Fermi vector
goes to zero at the same time as the Fermi velocity). 

The short dash, long dash-dot and short dash-dot lines on Fig.~\ref{Fig.2} show the boundaries
of these ``Fermi surface topology transitions''. Notably,
except for a range of densities near where the $l=1$ PI would have
been, the unpolarized Fermi sphere is only stable at small coupling

\section{Hard-core model}

\begin{figure}
\includegraphics[%
  bb=50bp 50bp 266bp 201bp,
  width=1.0\columnwidth]{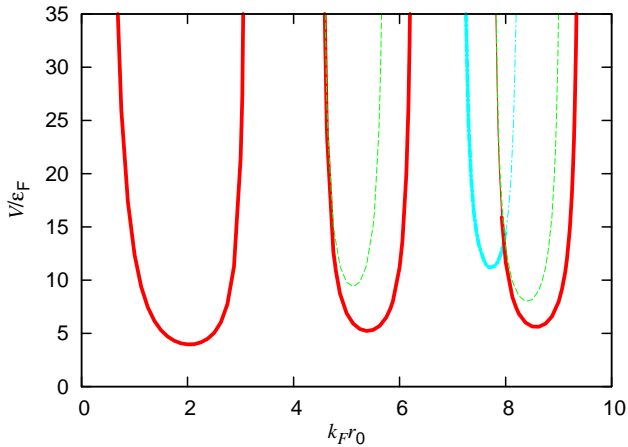}
\caption{\label{Fig.4} (color online) Phase diagram showing zero-temperature instabilities of the unpolarized, spherical Fermi surface for the ``hard-core'' model. The coupling constant is measured in units of the Fermi energy $\varepsilon_F=\hbar^2k_F^2/2m$. Key as in Fig.\ref{Fig.2}.}
\end{figure}

Our results suggest that the essential ingredient of the PI in the present
context, namely the finite range $r_{0}\sim k_{F}^{-1}$, also leads to the
interaction-driven Lifshitz transition. To probe the generality of this
observation, we have repeated the calculation (this time by evaluating the mean-field
equations (\ref{2.2}) and (\ref{2.4}) numerically) for a repulsive {}``hard core'' potential
\begin{equation}
V\left(\left|\mathbf{r}\right|\right)=V\Theta\left(r_{0}-\left|\mathbf{r}\right|\right).
\label{ficr}
\end{equation}
We have again found that, for $r_{0}\gtrsim k_{F}^{-1},$ there are, in addition to
the Stoner instability, PI with $l=2,4,6,\ldots$  Moreover we also find an
interaction-driven Lifshitz transition which, for certain ranges of values of
$k_Fr_0$, takes place before the Stoner or PI set in. A phase diagram is
presented in Fig.~\ref{Fig.4}. 

Unlike the delta-shell potential, for the hard core potential the PI domes are contained
within the Stoner ones, i.e. the PI can only take place on a polarized
Fermi surface, or for spinless fermions. However note that there are regions of the phase diagram where the boundary of the $l=2$ instability nearly overlaps with that for $l=0$, indicating that the two transitions happen almost simultaneously. 

The plot does not show all the instabilities: the $l>2$ PI take place at higher values of $k_Fr_0$ than those shown. The are also other domes of topological instability, though for this potential all of them are of the type in Fig.1~(e).

Our results for the hard-core potential not only support our identification of a sharp feature of $V\left(r\right)$ at $r_{0}\sim k_{F}^{-1}$ as the crucial ingredient for a PI, but suggest as well that, in such situations, the interaction-driven Lifshitz transition is at least as likely to occur as the Stoner or PI. 

We conclude this section by noting that a similar analysis using the screened Coulomb interaction
\begin{equation}
   V\left(\left|\mathbf{r}\right|\right)=\frac{e^2}{4\pi \epsilon_0^2}\frac{e^{-\left|\mathbf{r}\right|/r_0}}{\left|\mathbf{r}\right|}
   \label{scco}
\end{equation}
does not reveal either PI or instabilities of the Fermi surface topology ---only the Stoner instability is realized, and that only if we allow $r_0$ to deviate from its Thomas-Fermi value. 

\section{Conclusions}

In summary we have studied, at the mean field level, PI of a uniform
Fermi liquid with \emph{central} fermion-fermion interactions. We
find that PI of different symmetries may emerge from repulsive interactions
of \emph{sufficiently long, but finite} range $r_{0}\gtrsim k_{F}^{-1}$
(with the interesting exception of the $l=1$ PI which never takes
place). We have confirmed this by solving the theory analytically
for an explicit form of the interaction potential featuring repulsion at a particular distance $r_0$. Surprisingly we
have found that, in addition to the PI, there is also a new type of
Fermi surface instability: the interaction-driven Lifshitz transition.
This topological phase transition is even subtler than the PI and
seems to be generically associated with the class of models leading
to the PI. Further support for this picture is provided by analysis of an additional model, featuring hard-core repulsion. On the other hand the screened Coulomb interaction does not lead to these effects suggesting that a sharp feature (either a spike of repulsion or a sudden drop) must be present at the distance $r_0$.

Unlike the Lifshitz transition, the new quantum phase transition that we have described is fundamentally driven by interactions. Thus one would expect it to have a stronger thermodynamic signature. It will be of great interest, in the near future, to establish this signature and the properties of the novel state of matter this phase transition may lead to. 

\begin{acknowledgments}
We acknowledge discussions with J.M.F.~Gunn, C.~Hooley, M.~Haque,
B.~J.~Powell, S.~Simon, J.F.~Chalker, N.I.~Gidopoulos, M.W.~Long, G.~Volovik, A.H.~Castro-Neto, E.~Fradkin, S. Ramos and W.J.L. Buyers.
JQ thanks the University of Birmingham for hospitality and acknowledges
financial support by the Leverhulme Trust and an Atlas fellowship
awarded by CCLRC in association with St. Catherine's College, Oxford.
\end{acknowledgments}


\begin{thebibliography}{10}
\bibitem{tallon_99}J. L. Tallon \emph{et al.}, \emph{Phys. Stat. Sol. (b)} \textbf{215},
531 (1999).
\bibitem{kim_03}T. T. M Palstra \emph{et al.}, \emph{Phys. Rev. Lett.} \textbf{55},
2727 (1985); M. B. Walker \emph{et al.}, \emph{Phys. Rev. Lett.} \textbf{71},
002630 (1993); \textbf{}K. H. Kim \emph{et al.}, \emph{Phys. Rev.
Lett.} \textbf{91}, 256401 (2003).
\bibitem{grigera_04}S. A. Grigera \emph{et al.}, \emph{Science} \textbf{306}, 1154 (2004).
\bibitem{balicas_01}L. Balicas \emph{et al., Phys. Rev. Lett.} \textbf{87}, 067002 (2001).
\bibitem{annett_05}J. F. Annett, M. Krawiec and B. L. Gyorffy, cond-mat/0510591.
\bibitem{FFLO_in_atoms}T. Mizushima \emph{et al.}, \emph{Phys. Rev. Lett.} \textbf{94}, 060404
(2005).
\bibitem{FFLO_in_atoms-2}Kun Yang, \emph{Phys. Rev. Lett.} \textbf{95}, 218903 (2005).
\bibitem{goral_02}K. G{\'o}ral, L. Santos and M. Lewenstein, \emph{Phys. Rev. Lett.}
\textbf{88}, 170406 (2002).
\bibitem{DDW}S. Chakravarty, R. B. Laughlin, D. K. Morr and C. Nayak, \emph{Phys.
Rev. B} \textbf{63}, 094503 (2001).
\bibitem{DDW-2}U. Schollow{\"o}ck et al., Phys. Rev. Lett. \textbf{90}, 186401 (2003).
\bibitem{DDW-3}S. Lee, J. B. Marston and J. O. Fjaerestad, \emph{Phys. Rev. B} \textbf{72},
075126 (2005).
\bibitem{pomeranchuk_58}I. Ia. Pomeranchuk, \emph{JETP} \textbf{35}, 524-525 (1958).
\bibitem{QH}E. Fradkin and S. A. Kivelson, \emph{Phys. Rev. B} \textbf{59}, 8065
(1999).
\bibitem{QH-2}E. Fradkin, S. A. Kivelson, E. Manousakis and K. Nho, \emph{Phys.
Rev. Lett.} \textbf{84}, 1982 (2000). \textbf{}
\bibitem{2005-Varma-Zhu}C. M. Varma and Lijun Zhu, cond-mat/0502344.
\bibitem{HM}C. J. Halbloth and W. Metzner, \emph{Phys. Rev. Lett.} \textbf{85},
5162 (2000).
\bibitem{HM-2}V. Hankevych, I. Grote and F. Wegner, \emph{Phys. Rev. B} \textbf{66},
094516 (2002).
\bibitem{HM-3}W. Metzner, D. Rohe and S. Andergassen, \emph{Phys. Rev. Lett.} \textbf{91},
066402 (2003).
\bibitem{HM-4}A. Neumayr and W. Metzner, \emph{Phys. Rev. B} \textbf{67}, 035112
(2003).
\bibitem{HM-5}E. C. Carter and A. J. Schofield, \emph{Phys. Rev. B} \textbf{70},
045107 (2004). 
\bibitem{HM-6}The PI may also be favored by non-linearities in the electron dispersion
relation: D. G. Barci and L. E. Oxman, \emph{Phys. Rev. B} \textbf{67},
205108 (2003).
\bibitem{2004-Kivelson-Fradkin-Geballe}S. A. Kivelson, E. Fradkin and 
T. H. Geballe, \emph{Phys. Rev. B} \textbf{69}, 144505 (2004).
\bibitem{phase_diagram}Hae-Young Kee and Yong Baek Kim, \emph{Phys. Rev. B} \textbf{71},
184402 (2005).
\bibitem{phase_diagram-2}I. Khavkine, Chung-Hou Chung, V. Oganesyan and Hae-Young Kee, \emph{Phys.
Rev. B} \textbf{70}, 155110 (2004).
\bibitem{phase_diagram-3}H. Yamase, V. Oganesyan and W. Metzner, \emph{Phys. Rev. B} \textbf{72},
35114 (2005).
\bibitem{collective}V. Oganesyan, S. A. Kivelson and E. Fradkin, \emph{Phys. Rev. B} \textbf{64},
195109 (2001).
\bibitem{collective-2}Hae-Young Kee, \emph{Phys. Rev. B} \textbf{67}, 073105 (2003).
\bibitem{collective-3}J. Nilsson and A. H. Castro Neto, \emph{Phys. Rev. B} {\bf 72}, 195104 (2005).
\bibitem{collective-4}L. Dell'Anna and W. Metzner, cond-mat/0507532.
\bibitem{collective-5}M. J. Lawler \emph{et al.}, cond-mat/0508747.
\bibitem{varma_05}C. M. Varma, \emph{Phil. Mag.} \textbf{85}, 1657-1666 (2005).
\bibitem{leggett_75}A. J. Leggett, \emph{Rev. Mod. Phys.} \textbf{47}, 331 (1975).
\bibitem{liquid_crystal}S. A. Kivelson, E. Fradkin and V. J. Emery, 
\emph{Nature} {\bf 393}, 550 (1998).
\bibitem{nozieres_85}P. Nozi\`eres and S. Schmitt-Rink, \emph{J. Low. Temp. Phys.} \textbf{59},
195 (1985).
\bibitem{nozieres_92}P. Nozieres, \emph{J. Phys. I France} \textbf{2}, 443-458 (1992).
\bibitem{2004-Wu-Zhang}Congjun Wu and Shou-Cheng Zhang, {\it Phys. Rev. Lett.}
{\bf 93}, 36403 (2004).
\bibitem{2000-Quintanilla-Gyorffy}J. Quintanilla and B. L. Gyorffy,
\emph{Physica B} \textbf{284-288}, 421-422 (2000).
\bibitem{quintanilla_02}J. Quintanilla, B. L. Gyorffy, J. F. Annett and J. P. Wallington, \emph{Phys.
Rev. B} \textbf{66}, 214526 (2002).
\bibitem{lifshitz}I. M. Lifshitz, \emph{Zh. Eksp. Teor. Fiz.} \textbf{38}, 1569 (1960)
{[}\emph{Sov. Phys. JETP} \textbf{11}, 1130 (1960){]}.
\bibitem{lifshitz-2}A. A. Abrikosov, {}``Fundamentals of the Theory of Metals'' (Elsevier,
New York, 1988).
\bibitem{2006-Volovik}G. Volovik, cond-mat/0601103.
\bibitem{1993-MacDonald-Yang-Johnson}A. H. MacDonald, S. R. Eric Yang and M. D. Johnson, {\it Aust. J. Phys.} {\bf 46}, 345 (1993).
\bibitem{1994-Chamon-Wen}C. de C. Chamon and X. G. Wen, {\it Phys. Rev. B} {\bf 49}, 8227 ( 1994).
\bibitem{2002-Wan-Yang-Rezayi}Xin Wan, Kun Yang and E. H. Rezayi, {\it Phys. Rev. Lett.} {\bf 88}, 056802 (2002).
\bibitem{2006-CastroNeto-Guinea-Peres} A. H. Castro Neto, F. Guinea and N. M. R. Peres, {\it Phys. Rev. B} {\bf 73}, 205408  (2006).
\bibitem{2006-Yang-Sachdev}Kun Yang and Subir Sachdev, {\it Phys. Rev. Lett.} {\bf 96}, 187001 (2006).
\bibitem{varma1}C.M. Varma, {\it Phys. Rev. B} {\bf 55}, 14554 (1997); {\it Phys. Rev. Lett.} {\bf 83},
	3538 (1999).
\bibitem{varma2}B. Fauque, Y. Sidis, V. Hinkov, S. Pailhes, C. T. Lin, X.
	Chaud, and P. Bourges, {\it Phys. Rev. Lett.} {\bf 96}, 197001 (2006).
\end{thebibliography}
\end{document}